\documentclass[conference]{IEEEtran}
\usepackage{amssymb,amsmath,amsfonts}
\usepackage{color}
\usepackage{mathtools}
\usepackage{mathrsfs}
\usepackage{bm}
\usepackage{graphicx}		
\usepackage{epstopdf}		
\usepackage[font={small}]{caption}
\usepackage{caption}
\usepackage{subcaption}
\usepackage{multicol}
\usepackage{enumerate}
\usepackage{units}
\usepackage[numbers]{natbib}
\usepackage{algorithm}

\usepackage{flushend}
\DeclarePairedDelimiter\abs{\lvert}{\rvert}

\usepackage{lipsum}

\newcommand\blfootnote[1]{%
  \begingroup
  \renewcommand\thefootnote{}\footnote{#1}%
  \addtocounter{footnote}{-1}%
  \endgroup
}

\begin{document}

\title{Spectral Efficiency of Multi-User Adaptive Cognitive Radio Networks}

%\author{H. Saki, \IEEEmembership{Member,~IEEE}, A. Shojaeifard, \IEEEmembership{Member,~IEEE}, \\ M.Shikh Bahaei, \IEEEmembership{Senior Member,~IEEE}
%
%\thanks{\scriptsize{H. Saki, A. Shojaeifard, M. Shikh-Bahaei. M.G. Martini are with the Institute of Telecommunications (IoT), King's College London, Strand, London, WC2R 2LS, UK. Tel: +44 (0)20 7848 1857, Fax: +44 (0)20 7848 2932. Email: \{h.saki,a.shojaeifard,m.sbahaei\}@kcl.ac.uk}.}
%}
%
%\markboth{Submitted to IEEE Transactions on Vehicular Technology, Jun. 2017}{SAKI \MakeLowercase{\textit{et al.}}: Spectral Efficiency of Adaptive MQAM/OFDMA Cognitive Radio Networks}
%
%\maketitle
%\setcounter{page}{1}

\author{\IEEEauthorblockN{H. Saki*, M.Shikh Bahaei*}
\IEEEauthorblockA{*Centre for Telecommunications Research (CTR), King's College London, UK\\
E-mail: \ hadi.saki@kcl.ac.uk, m.sbahaei@kcl.ac.uk}}

%\author{H. Saki, \IEEEmembership{Member,~IEEE},  \linebreak M. G. Martini, \IEEEmembership{Senior Member,~IEEE}

%\thanks{\scriptsize{H. Saki, and M. G. Martini are with the Kingston University London, Penrhyn Road, Kingston upon Thames, London, UK, KT1 2EE. Tel: +44 (0)20 20 8417 2699, Fax: +44 (0)20 8417 2972. Email: \{h.saki,m.nasralla,m.martini\}@kingston.ac.uk}}}
%
%\markboth{Draft Jan. 2017}{Saki \MakeLowercase{\textit{et al.}}: Video Traffic Classification for Uplink Video Transmission over LTE Networks}
\maketitle
\setcounter{page}{1}

\begin{abstract}
%In this correspondence, optimal radio resource allocation algorithms facilitating power, rate, and subcarrier adaptation have been investigated for enhancing the spectral efficiency of multi-user orthogonal frequency-division multiple access (OFDMA) cognitive radio (CR) networks subject to satisfying power and interference constraints. Novel algorithms have been proposed under different scenarios with deterministic and probabilistic interference violation limits based on perfect and imperfect availability of cross-link channel state information (CSI).  
In this correspondence, the comprehensive problem of joint power, rate, and subcarrier allocation have been investigated for enhancing the spectral efficiency of multi-user orthogonal frequency-division multiple access (OFDMA) cognitive radio (CR) networks subject to satisfying total average transmission power and aggregate interference constraints. We propose novel optimal radio resource allocation (RRA) algorithms under different scenarios with deterministic and probabilistic interference violation limits based on perfect and imperfect availability of cross-link channel state information (CSI). In particular, we propose a probabilistic approach to mitigate the total imposed interference on the primary service under imperfect cross-link CSI. A close form mathematical formulation of the cumulative density function (cdf) for the received signal-to-interference-plus-noise ratio (SINR) is formulated to evaluate the resultant average spectral efficiency (ASE). Dual decomposition is utilized to obtain sub-optimal solutions for the non-convex optimization problems. Through simulation results, we investigate the achievable performance and the impact of parameters uncertainty on the overall system performance.  Furthermore, we present that the developed RRA algorithms can considerably improve the cognitive performance whilst abiding to the imposed power constraints. In particular, the performance under imperfect cross-link CSI knowledge for the proposed `probabilistic case' is compared over the conventional scenarios to show the potential gain in employing this scheme.
\end{abstract}

%\begin{keywords}
%Cognitive radio, OFDMA, RRA, interference management, posteriori probability estimation, cross-link channel state information, probabilistic analysis, spectral efficiency.
%\end{keywords}
\blfootnote{*This work was partially supported by the UK Engineering and Physical Sciences Research Council (EPSRC) grant numbers EP/P022723/1 and EP/P003486/1.}
\section{Introduction}
In recent years, a significant effort has been made towards improving the spectral efficiency of cellular networks in order to meet the growing demand and sophistication of wireless applications. Several technologies, such as small-cell (SC) solution, Full Duplex (FD) \cite{tow} ,cognitive radio (CR) networks and massive multiple-input multiple-output (MIMO),  \cite{7933143}- each with respective advantages and challenges - are promising candidates in this direction \cite{6191306}. CR is outlined as an smart radio network that has the ability to sense the primary service behaviour and surrounding environment and adjust its spectrum usage and parameters based on the observed information \cite{788210}. 
According to Ofcom flexible spectrum-sharing for supporting CR and spectrum co-existence is a priority issue to overcome the current capacity crunch.
% and thus enabling the deployment of long term evolution (LTE)-advanced and beyond wireless systems.  

%Three main scenarios have been proposed for cognitive radio in regards to the unlicensed users access to the primary frequency band: (i) underlay spectrum access where secondary users silently coexist with primary users, provided they satisfy an interference limit set by a regulatory authority (ii) overlay spectrum access in which secondary users are only allowed to access the vacant parts of the primary spectrum, and (iii) hybrid spectrum access, a combination of the two former strategies in which the secondary users sense the primary spectrum and adjust their transmission parameters based on the detection, whilst avoid imposing harmful interference to the primary users. 
%In this paper, we consider underlay spectrum-sharing, where robust interference management is critical for tackling any harmful cross-service interference. 

Orthogonal frequency-division multiplexing (OFDM) has emerged as a prominent technology for new generation of wireless communication systems and adopted in many modern wireless technologies. For OFDM-based multi-user applications, multiple-access can be accommodated through orthogonal frequency-division multiple-access (OFDMA) technique. In OFDMA systems, different subcarriers may be assigned to different users in order to exploit the random variations of users across each subcarrier. OFDMA is considered as a effective technology for CR networks as a result of the inherent advantages with regards to adaptability and flexibility in allocating radio resources in a shared-spectrum environments.

Radio resource Management (RRM) plays an important role in optimizing the spectral efficiency of conventional OFDMA systems \cite{7924379,5506103,4786499}. Adaptive RRM is a prominent field for research in the context of multi-user CR networks with the aim of obtaining a balance between the CR performance and reducing the induced interference on the primary users. Optimal and suboptimal RRM policies are studied in \cite{7881785}, where the aggregate throughput of the CR system is maximized under a primary receiver (PRx) interference limit. Dynamics and adaptive MAC retry-limit aware adaptation is proposed in \cite{4939126}. A queue-aware RRA algorithm is proposed in \cite{5557657}, a to maximize the fairness in OFDMA-based CR networks subject to a total power constraint at the base station. A Lagrangian relaxation algorithm is adopted in \cite{5510778} to probabilistically allocate resources based on the availability of the primary frequency band via spectrum sensing. 

Most of the RRA algorithms for CR networks in the literature assume perfect channel state information (CSI) between the cognitive transmitter (CTx) and PRx, and few have considered imperfect cross-link CSI. However, due to technical reasons such as estimation errors and wireless channel delay, obtaining perfect cross-link CSI is difficult in practical scenarios.
In \cite{5419086} and \cite{6185693}, the ergodic capacity is derived over fading channels with imperfect cross-link knowledge, however, the analysis is carried out for a single cognitive user (CU). Furthermore, due to noisy cross-link information, it is unrealistic to assume that the secondary network strictly satisfies a deterministic interference constraint. The authors in \cite{5967979} propose a RRA algorithm for maximizing instantaneous rate in downlink OFDMA CR systems subject to satisfying a collision probability constraint. However, \cite{5967979} only considers the individual impact of probabilistic interference constraint per subcarrier. 
%Motivated by the above, we thoroughly investigate different scenarios by analysing the impact of deterministic and probabilistic interference constraints depending on perfect and noisy cross-link knowledge.

To the best of authors' knowledge, enhancing the average spectral efficiency of multi-user OFDMA-based CR systems has not been addressed in the literature. In this work, by exploiting the advantages of channel adaptation techniques, we propose novel joint power, subcarrier, and rate allocation algorithms for enhancing the average spectral efficiency of downlink multi-user adaptive M-ary quadrature amplitude modulation (MQAM)/OFDMA CR systems. Given the received power restrictions on the CTx in order to satisfy the primary network interference limit and the cognitive network power constraint, the CTx transmit power is a function of the cognitive-cognitive direct-link and cognitive-primary cross-link fading states. We develop a closed-form expression for the cumulative distribution function (cdf) of the CR's received signal-to-interference-plus-noise ratio (SINR) to evaluate the average spectral efficiency of the adaptive MQAM/OFDMA CR system.  
      
%The main novelties and contributions of this paper are summarized as follows:
%\begin{enumerate}
%\item The comprehensive problem of power, rate, and subcarrier allocation for enhancing the average spectral efficiency of downlink multi-user OFDMA CR systems subject to satisfying total average transmission power and peak aggregate interference constraint has been studied.
%
%\item A closed-form expression for the cdf of the OFDMA CR's received SINR is derived under limitations imposed on the CTx through the power and interference constraints. Consequently, an upper-bound expression for average spectral efficiency of the adaptive multi-user MQAM/OFDMA CR system is formulated.
%
%\item The critical issue of violating interference limits associated with imperfect cross-link CSI availability is examined by carrying out the analysis for the `probabilistic case' scenario of channel estimation error. 
%%
%%\item The impact of deterministic and probabilistic interference constraints on the system performance is considered with perfect and imperfect cross-link CSI. In particular, we propose a new low-complexity deterministic formulation for the probabilistic cross-link interference. 
%\end{enumerate}

The organization of this paper is as follows: Section II presents the network model and operation assumptions. In Section III, the resource allocation problem for enhancing average spectral efficiency of the adaptive multi-user MQAM/OFDMA under perfect cross-link CSI subject to power and deterministic interference constraints is developed. Section IV investigates the performance under a collision probability constraint with imperfect cross-link CSI and proposes a deterministic formulation of the probabilistic aggregate cross-link interference. Illustrative numerical results for various scenarios under consideration are provided in Section V. Finally, concluding remarks are presented in Section VI. 

\section{System Model and Preliminaries}

In this section, the multi-user OFDMA CR network model and operation assumptions are introduced. Further, interference management schemes and spectral efficiency of the adaptive MQAM/OFDMA system under consideration are studied.   

\subsection{Network Architecture and Wireless Channel}

\begin{figure}[t]
\centering
\includegraphics[width=.375\textwidth]{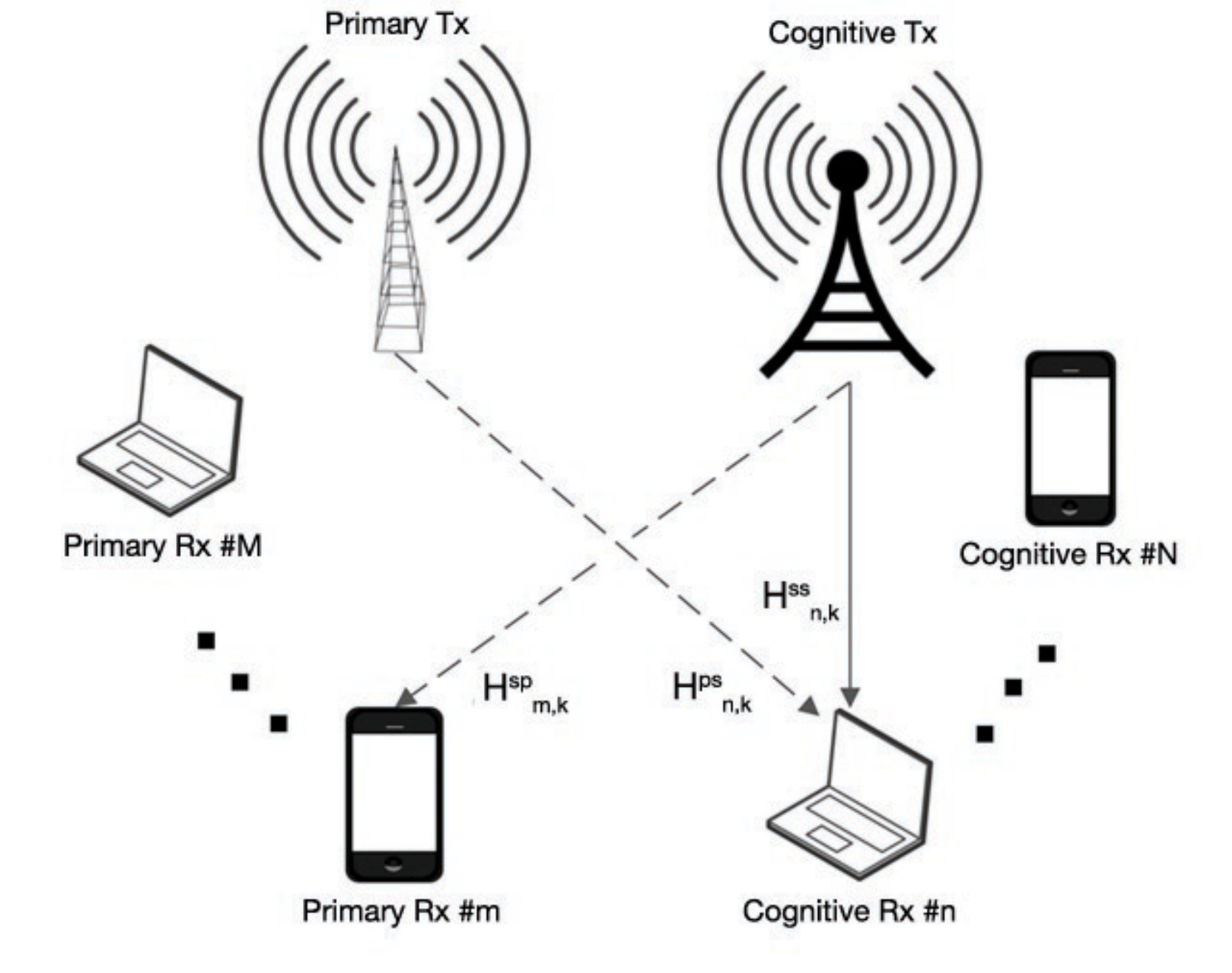}
\caption{Schematic diagram of the shared-spectrum OFDMA system. For simplicity purposes, channels of a single cognitive user are drawn.}
\label{CRmodel}
\end{figure}

We consider an underlay shared-spectrum environment, as shown in Fig. \ref{CRmodel}, where a cognitive network with a single CTx and $n \in \lbrace 1, ..., N \rbrace$ cognitive receivers (CRx)s coexist with a primary network with a primary transmitter (PTx) and $m \in \lbrace 1, ..., M \rbrace$ PRxs. The cognitive network can access a spectrum licensed to the primary network with a total bandwidth of $B$ which is divided into $K$ non-overlapping sub-channels subject to not violating the imposed interference constraint set by a regulatory authority. The sub-channel bandwidth is assumed to be much smaller than the coherence bandwidth of the wireless channel, thus, each subcarrier experiences frequency-flat fading. Let $H^{ss}_{n,k}(t)$, $H^{ps}_{n,k}(t)$, and $H^{sp}_{m,k}(t)$, at time $t$, denote the channel amplitude gains over subchannel $k$ from the CTx to $n^{\text{th}}$ CRx, PTx to the $n^{\text{th}}$ CRx, and CTx to $m^{\text{th}}$ PRx respectively. The channel power gains $|H^{ss}_{n,k}(t)|^2$, $|H^{ps}_{n,k}(t)|^2$, and $|H^{sp}_{m,k}(t)|^2$ are assumed to be ergodic and stationary with continuous probability density functions (pdf)s $f_{|H^{ss}_{n,k}(t)|^2}(.)$, $f_{|H^{ps}_{n,k}(t)|^2}(.)$, and $f_{|H^{sp}_{m,k}(t)|^2}(.)$, respectively. In addition, the estimated instantaneous values and distribution information of the secondary-secondary channel power gains is assumed to be available at the CTx \cite{6185693}. In this work, we consider different cases with perfect and noisy cross-link knowledge between CTx and PRxs. 

Each sub-channel is assigned exclusively to at most one CRx at any given time, hence, there is no mutual interference between different cognitive users \cite{6104418}. It should also be noted that by utilizing an appropriate cyclic prefix, the inter-symbol-interference (ICI) can be ignored \cite{5672622}. The received SINR of cognitive user $n$ over sub-channel $k$ at time instant $t$ is 
\begin{align}
\gamma_{n,k}(t) = \frac{P_{n,k} |H^{ss}_{n,k}(t)|^2}{\sigma^2_{n} + \sigma^2_{ps}}
\end{align}
where $P_{n,k}$ is a fixed transmit power allocated to cognitive user $n$ over sub-channel $k$, $\sigma^{2}_{n}$ is the noise power, and $\sigma^{2}_{ps}$ is the received power from the primary network. Without loss of generality, $\sigma^{2}_{n}$ and $\sigma^{2}_{ps}$ are assumed to be the same across all users and sub-channels \cite{5290301,5672619}. Assuming stationariness of the channel gain, for the sake of brevity, we henceforth omit the time reference $t$.
% We define $\Upsilon_{n,k}(t)$ as a vector containing $\gamma_{n,k}(t)$ of all time intervals.     

Due to the impact of several factors, such as channel estimation error, feedback delay, and mobility, perfect cross-link information is not available \cite{kob}. With noisy cross-link CTx to PRxs knowledge, we model the inherent uncertainty in channel estimation in the following form
\begin{align}
H^{sp}_{m,k} = \hat{H}^{sp}_{m,k} + \Delta H^{sp}_{m,k}
\label{errorF}
\end{align}
where over subcarrier $k$, $H^{sp}_{m,k}$ is the actual cross-link gain, $\hat{H}^{sp}_{m,k}$ is the channel estimation considered to be known, and $\Delta H^{sp}_{m,k}$ denotes the estimation error. $H^{sp}_{m,k}$, $\hat{H}^{sp}_{m,k}$, and $\Delta H^{sp}_{m,k}$ are assumed to be zero-mean complex Gaussian random variables with respective variances $\delta^{2}_{H^{sp}_{m,k}}$, $\delta^{2}_{\hat{H}^{sp}_{m,k}}$, and $\delta^{2}_{\Delta H^{sp}_{m,k}}$ \cite{5419086,4801449}. For robust receiver design, we consider the estimation $\hat{H}^{sp}_{m,k}$ and error $\Delta H^{sp}_{m,k}$ to be statistically correlated random variables with a correlation factor $\rho = \sqrt{\delta^{2}_{\Delta H^{sp}_{m,k}}/(\delta^{2}_{\Delta H^{sp}_{m,k}} + \delta^{2}_{H^{sp}_{m,k}})}$, where $0 \leq \rho \leq 1$. 

\subsection{Interference Management}

In a shared-spectrum environment, and particularly for delay-sensitive services, the licensed users' quality of service (QoS) is highly dependent on the instantaneous received SINRs of cognitive users. In order to protect the licensed spectrum from harmful interference we impose a deterministic peak total interference constraint between CTx and primary users
\begin{align}
\sum_{n=1}^{N} \sum_{k=1}^{K} \varphi_{n,k}(\gamma_{n,k}) P_{n,k}(\gamma_{n,k}) |H^{sp}_{m,k}|^2 \leq I^{m}_{th} , \forall m \in \{1, ..., M\}
\label{IMeq1}
\end{align}
%where $\Upsilon$ is a matrix containing all of $\Upsilon_{n,k}$,  $\forall n \in \{1, ..., N\}$ and $\forall k \in \{1, ..., K\}$,
where $\varphi_{n,k}(\gamma_{n,k})$ is the time-sharing factor (subcarrier allocation policy), $P_{n,k}(\gamma_{n,k})$ is the allocated transmit power, and $I^{m}_{th}$ denotes the maximum tolerable interference threshold. 

However, as a consequence of uncertainties about the shared-spectrum environment and primary service operation, it is unrealistic to assume that the CTx always satisfies the deterministic peak total interference constraint. In practical scenarios, probability of violating the interference is confined to a certain value that satisfies the minimum QoS requirements of primary users. Probabilistic interference constraint is particularly critical for robust interference management given noisy cross-link knowledge. To improve the overall system performance and to mitigate the
impact of channel estimation errors, the following allowable probabilistic interference limit violation is considered
\begin{gather}
\mathscr{P} \left( \sum_{n=1}^{N} \sum_{k=1}^{K} \varphi_{n,k}(\gamma_{n,k}) P_{n,k}(\gamma_{n,k}) |H^{sp}_{m,k}|^2 > I^{m}_{th} \right) \leq \epsilon^{m} \nonumber \\ , \forall m \in \{1, ..., M\}
\label{IMeq2}
\end{gather}
where $\mathscr{P}(.)$ denotes probability, and $\epsilon^{m}$ is the collision probability constraint for the $m^{\text{th}}$ PRx. 
%Note that the constraints in (\ref{IMeq1}) and (\ref{IMeq2}) are defined under perfect cross-link CSI knowledge.  

On the other hand, mitigating the interference between neighbouring cells is a vital issue due to the increasing frequency reuse in modern wireless communication systems \cite{5770666}. As a remedy to inter-cell interference, and to maintain effective and efficient power consumption, we impose a total average transmit power constraint on the cognitive network as follows 
\begin{align}
\sum_{n=1}^{N} \sum_{k=1}^{K} E_{\gamma_{n,k}} \Big\{ \varphi_{n,k}(\gamma_{n,k}) P_{n,k}(\gamma_{n,k}) \Big\} \leq P_t.
\end{align}
where $E_{x}(.)$ denotes the expectation with respect to $x$, and $P_{t}$ denotes the total average transmit power limit. 

\subsection{Spectral Efficiency}

The focus of this work is mainly on optimal power, rate, and subcarrier allocation for enhancing the average spectral efficiency of the adaptive MQAM/OFDMA CR network. In a multi-user scenario, various subcarriers may be allocated to different users. In other words, users may experience different channel fading conditions over each sub-channel. Therefore, any efficient resource allocation scheme in OFDMA must be based on the sub-channel quality of each user. Furthermore, in a shared-spectrum environment, satisfying the interference constraints is an important factor in allocating resources.   

Employing square MQAM with Gray-coded bit mapping, the approximate instantaneous bit-error-rate (BER) expression for user $n$ over subcarrier $k$ is given by \cite{5672622,4917798},
\begin{align}
\xi^{b}_{n,k}(\gamma_{n,k}) & = \frac{4}{\log_{2}(M_{n,k}(\gamma_{n,k}))} \Biggr(1 - \frac{1}{\sqrt{M_{n,k}(\gamma_{n,k})}} \Biggr) \nonumber \\ & \times Q \Biggr(\sqrt{\frac{3 \gamma_{n,k}}{M_{n,k}(\gamma_{n,k})-1}} \Biggr)
\label{approxBER}
\end{align} 
where $M_{n,k}(\gamma_{n,k})$ denotes the constellation size vector of MQAM when each element is a function of the instantaneous received SINR of the cognitive user $n$ over subcarrier $k$, and $Q(.)$ represents the Gaussian Q-function. 

The average spectral efficiency of the adaptive multi-user MQAM/OFDMA system per subcarrier per user over the fading channel is defined as 
\begin{align}
ASE = \sum_{n=1}^{N} \sum_{k=1}^{K} E_{\gamma_{n,k}} \Big\{ \log_{2} \left( M_{n,k}(\gamma_{n,k}) \varphi_{n,k}(\gamma_{n,k}) \right) \Big\}.
\label{AASE}
\end{align} 
In order to evaluate the $ASE$, the distribution of the received SINR, a function of secondary-secondary and secondary-primary channels, must be developed. 

\section{Deterministic Interference Constraint with Perfect Cross-Link CSI}

%The objective of this paper is to maximize the aggregate average spectral efficiency of cognitive users while satisfying total transmission power and peak maximum tolerable interference constraints.
 In this section, we solve the resource allocation problem with perfect cross-link knowledge and deterministic interference constraint.

\subsection{Problem Formulation}
Mathematically, the optimization problem can be stated as follows.

\textit{Problem} $\mathscr{O}_{1}$:
\begin{subequations}
\label{optimizationProb}
\begin{gather}
\max_{\varphi_{n,k},P_{n,k}} \sum_{n=1}^{N} \sum_{k=1}^{K} E_{\gamma_{n,k}} \Big\{ \log_{2}(M_{n,k}(\gamma_{n,k})) \varphi_{n,k}(\gamma_{n,k}) \Big\}  
\label{OF1a}\\
\text{s. t.:} \quad \sum_{n=1}^{N} \sum_{k=1}^{K} E_{\gamma_{n,k}} \Big\{  \varphi_{n,k}(\gamma_{n,k}) P_{n,k}(\gamma_{n,k}) \Big\} \leq P_t \label{OF1b}\\
\sum_{n=1}^{N} \sum_{k=1}^{K} \varphi_{n,k}(\gamma_{n,k}) P_{n,k}(\gamma_{n,k}) |H^{sp}_{m,k}|^2 \! \leq \! I^{m}_{th} , \forall m \! \in \! \{1, ..., M\} \label{OF1c}\\
\sum_{n=1}^{N} \varphi_{n,k}(\gamma_{n,k}) = 1 , \forall k \! \in \! \{1, ..., K\} \label{OF1d}\\
\varphi_{n,k}(\gamma_{n,k}) \in \{0,1\} , \forall n \in \{1, ..., N\} , \forall k \in \{1, ..., K\}  \label{OF1be}\\
\xi^{b}_{n,k}(\gamma_{n,k}) \leq \xi , \forall n \in \{1, ..., N\} , \forall k \in \{1, ..., K\} \label{OF1e}
\end{gather}
\end{subequations}
%where $p_{\gamma[.]}(\gamma[.])$ is the pdf of $\gamma[.]$, 
where $\xi$ denotes the common BER-target.

In the adaptive multi-user MQAM/OFDMA \cite{4382916,neh} CR system under consideration, different transmit power and constellation sizes are allocated to different users and subcarriers. Using the upper-bound expression for the Gaussian Q-function, i.e., $Q(x) \leq (1/2) \exp(-x^2/2)$, the instantaneous BER for user $n$ over subcarrier $k$, subject to an instantaneous constraint $\xi^{b}_{n,k}(\gamma_{n,k}) = \xi$ can be expressed as 
\begin{align}
\xi^{b}_{n,k}(\gamma_{n,k}) \leq 0.3 \exp \Biggr( \frac{-1.5  \gamma_{n,k} }{M_{n,k}(\gamma_{n,k}) - 1 } \frac{P_{n,k}(\gamma_{n,k})}{\min \left(\frac{P_t}{K} , \frac{I^{m}_{th}}{N^{sp}_{m}} \right)} \Biggr). 
\end{align}
where $N^{sp}_{m} = \sum_{k=1}^{K} |H^{sp}_{m,k}|^2$. With further manipulation, for a BER-target $\xi$, the maximum constellation size for user $n$ over subcarrier $k$ is obtained as 
\begin{align}
M^{*}_{n,k}(\gamma_{n,k}) = 1 + \frac{\zeta   \gamma_{n,k} P_{n,k}(\gamma_{n,k}) }{\min \left(\frac{P_t}{K} , \frac{I^{m}_{th}}{N^{sp}_{m}} \right)} 
\label{optConstSize}
\end{align}
where 
\begin{align}
\zeta = \frac{-1.5}{\ln(\xi/0.3)}.
\end{align}

According to the constraints (\ref{OF1b}) and (\ref{OF1c}) in the optimization problem $\mathscr{O}_{1}$, the joint cumulative density function (cdf) of $\gamma_{n,k}$ can be writtenas:
\begin{align}
F_{\gamma_{n,k}}(\Gamma) = \mathscr{P} \biggr( \frac{P_{t} |H^{ss}_{n,k}|^2}{K (\sigma^2_{n} + \sigma^2_{ps})} \leq \Gamma , \frac{I^{m}_{th} |H^{ss}_{n,k}|^2}{N^{sp}_{m} (\sigma^2_{n} + \sigma^2_{ps})} \leq \Gamma \biggr).
\label{cdfofG}
\end{align}
The probability expression in (\ref{cdfofG}) can be further simplified by considering the cases $\frac{P_{t} |H^{ss}_{n,k}|^2}{K (\sigma^2_{n} + \sigma^2_{ps})} \lesseqqgtr \frac{I^{m}_{th} |H^{ss}_{n,k}|^2}{N^{sp}_{m} (\sigma^2_{n} + \sigma^2_{ps})}$ and conditioning on $N^{sp}_{m}$
\begin{align}
& 1 - \mathscr{P} \biggr( \frac{P_{t} |H^{ss}_{n,k}|^2}{K (\sigma^2_{n} + \sigma^2_{ps})} > \Gamma , \frac{I^{m}_{th} |H^{ss}_{n,k}|^2}{N^{sp}_{m} (\sigma^2_{n} + \sigma^2_{ps})} > \Gamma \biggr) = \nonumber \\ & 1 -
  \begin{dcases}
   \mathscr{P} \biggr( |H^{ss}_{n,k}|^2 > \frac{K \Gamma (\sigma^2_{n} + \sigma^2_{ps})}{P_{t}} \biggr) & N^{sp}_{m} \leq \frac{I^{m}_{th} K}{P_t} \nonumber \\
   \mathscr{P} \biggr( |H^{ss}_{n,k}|^2 > \frac{N^{sp}_{m} \Gamma (\sigma^2_{n} + \sigma^2_{ps})}{I^{m}_{th}} \biggr) & N^{sp}_{m} > \frac{I^{m}_{th} K}{P_t}.
  \end{dcases}\\
  \label{sigmaOPT}
\end{align}

\textit{Lemma 1:}
For large values of $K$, given complex Gaussian random variables $H^{sp}_{m,k}$ with means $\mu_{H^{sp}_{m,k}}$ and equal variance $\delta^{2}_{H^{sp}_{m,k}}$ for all $k \in \{1, ..., K\}$, the non-central Chi-square random variable $N^{sp}_{m} = \sum_{k=1}^{K} |H^{sp}_{m,k}|^2$ can be approximated as a Gaussian random variable with respective mean and variance $\mu_{N^{sp}_{m}} \! = \! \delta_{H^{sp}_{m,k}}^{2} \left[ 2K \! + \! \mu^{'} \right]$ and\linebreak $\delta^{2}_{N^{sp}_{m}} \! = \! \delta^4_{H^{sp}_{m,k}} \left[ 4K \! + \! 4 \mu^{'} \right]$, where ${\mu^{'}} = \sum_{k=1}^{K}(\frac{\mu_{H^{sp}_{m,k}}}{\delta_{H^{sp}_{m,k}}})^2$.\\

\textit{Proof:} We can write $H^{sp}_{m,k} = \delta_{H^{sp}_{m,k}} G^{sp}_{m,k}$, where $G^{sp}_{m,k} \thicksim CN(\frac{\mu_{H^{sp}_{m,k}}}{\delta_{H^{sp}_{m,k}}},1)$. Assuming equal variance for random variables $H^{sp}_{m,k}$, $\sum_{k=1}^{K} |G^{sp}_{m,k}|^2$ is a non-central Chi-Square random variable with  $2K$ degrees of freedom and non-centrality parameter $\mu^{'} = \sum_{k=1}^{K}(\frac{\mu_{H^{sp}_{m,k}}}{\delta_{H^{sp}_{m,k}}})^2$.
%Note that the distribution of $\sum_{k=1}^{K} |G^{sp}_{m,k}|^2$ depends only on the sum of $(\frac{\mu_{sp,1}}{\delta_{sp,1}})^2 + ... + (\frac{\mu_{sp,K}}{\delta_{sp,K}})^2$ and not on the individual values of $(\frac{\mu_{sp,k}}{\delta_{sp,k}})$. As a result of this, we can assume that $(\frac{\mu_{sp,1}}{\delta_{sp,1}}) = ... = (\frac{\mu_{sp,K-1}}{\delta_{sp,K-1}}) = 0$ and $(\frac{\mu_{sp,K}}{\delta_{sp,K}}) = \mu^{'}$. Then
%\begin{gather}
%\sum_{k=1}^{K} |G^{sp}_{m,k}|^2 = |S_{1}|^{2} + ... + |S_{K-1}|^2 + |S_{K} + \mu^{'}|^{2} \nonumber \\ = |S_{1}|^{2} + ... + |S_{K-1}|^{2} + |S_{K}|^{2}  + 2 |S_{K}| (\mu^{'}) + (\mu^{'})^{2} \nonumber \\
%\end{gather}
%where $S_{k} \thicksim CN(0,1)$ are independent complex Gaussian random variables. 
For large values of $K$, central limit theorem (CLT) can be invoked to show that the non-central Chi-Square random variable $\sum_{k=1}^{K} |G^{sp}_{m,k}|^2$ can be approximated as a Gaussian random variable as follows
\begin{align}
\sum_{k=1}^{K} |G^{sp}_{m,k}|^{2} \thicksim N \left( 2 K \! + \! \mu^{'},4K \! + \! 4 \mu^{'} \right).
\label{Gsp1}
\end{align} 
%Given $|S_{k}| \thicksim \text{Rayleigh}(1)$, due to the significance of random variable $\sum_{k=1}^{K} |S_{k}|^{2}$ in (\ref{Gsp1}) for large values of $K$, the following approximation holds with high accuracy
%\begin{align}
%2 |S_{K}| (\mu^{'}) \! + \! (\mu^{'})^{2} \! \thicksim \! N \biggr(\! \sqrt{2 \pi} \mu^{'} \! + \! (\mu^{'})^{2}, 2 (4 \! - \! \pi)(\mu^{'})^{2} \! \biggr).
%\label{RaytoNorm1}
%\end{align}
Hence, $N^{sp}_{m} = \sum_{k=1}^{K} |H^{sp}_{m,k}|^2$ can be approximated by
\begin{align}
N^{sp}_{m} = \sum_{k=1}^{K} |H^{sp}_{m,k}|^2 \thicksim N \biggr( \mu_{N^{sp}_{m}} , \delta^{2}_{N^{sp}_{m}} \biggr)
\label{NCCtoN}
\end{align}
where $\mu_{N^{sp}_{m}} = \delta^2_{H^{sp}_{m,k}} \Big[ 2K + \mu^{'} \Big]$ and $\delta^{2}_{N^{sp}_{m}} = \delta^{4}_{H^{sp}_{m,k}} \Big[ 4K + 4 \mu^{'} \Big]$. 
%In order to investigate the proposed approximation, we compare the cdfs of left hand side (LHS) and right hand side (RHS) of (\ref{NCCtoN}) in Fig. 2. 
Denoting the pdf of $N^{sp}_{m}$ with $f_{N^{sp}_{m}}(.)$, and the cdfs of $|H^{ss}_{n,k}|^2$ and $N^{sp}_{m}$ with $F_{|H^{ss}_{n,k}|^2}(.)$ and $F_{N^{sp}_{m}}(.)$, respectively, we obtain the cdf of $\gamma_{n,k}$ as
\begin{align}
F_{\gamma_{n,k}}(\Gamma) & = 1 -  A - B,
\end{align}
\begin{align}
& A = \!\! \int^{\frac{I^{m}_{th} K}{P_t}}_{0} \!\!\!\! \mathscr{P} \! \left(\! |H^{ss}_{n,k}|^2 \! > \! \frac{K \Gamma (\sigma^2_{n} + \sigma^2_{ps})}{P_{t}} \! \right) \! f_{N^{sp}_{m}}(N^{sp}_{m}) d N^{sp}_{m} \nonumber \\ &  = \! \mathscr{P} \! \left(\! |H^{ss}_{n,k}|^2 \! > \! \frac{K \Gamma (\sigma^2_{n} + \sigma^2_{ps})}{P_{t}} \! \right) \int^{\frac{I^{m}_{th} K}{P_t}}_{0} \!\!\!\! f_{N^{sp}_{m}}(N^{sp}_{m}) \, d N^{sp}_{m} \nonumber \\ & = \mathscr{P} \! \left(\! |H^{ss}_{n,k}|^2 \! > \! \frac{K \Gamma (\sigma^2_{n} + \sigma^2_{ps})}{P_{t}} \! \right) \mathscr{P} \left( N^{sp}_{m} \leq \frac{I^{m}_{th} K}{P_t} \right) \nonumber \\ & = \left( 1 - F_{|H_{ss}|^2} \biggr( \frac{K \Gamma (\sigma^2_{n} + \sigma^2_{ps})}{P_{t}} \biggr) \right) \, F_{N^{sp}_{m}} \left( \frac{I^{m}_{th} K}{P_t} \right) 
\label{integA}
\end{align}
and
\begin{align}
\! B \! = \!\!\! \int_{\frac{I^{m}_{th} K}{P_t}}^{\infty} \!\! \mathscr{P} \! \biggr(\! |H^{ss}_{n,k}|^2 \!\! > \!\! \frac{N^{sp}_{m} \Gamma (\sigma^{2}_{n} \! + \! \sigma^{2}_{ps})}{I^{m}_{th}} \! \biggr) f_{N^{sp}_{m}}(N^{sp}_{m}) \, d N^{sp}_{m}.
\label{integB}
\end{align}
Recall that the cdf of a Normally-distributed random variable $X$ with mean $\mu$ and standard deviation $\sigma$ is given by $F_{X}(x)=\frac{1}{2} \left[ 1 + erf \left( \frac{x - \mu}{2 \sigma^2} \right) \right]$, and the cdf of an Exponentially-distributed random variable $Y$ is computed by $F_{Y}(y) = 1 - e^{-y/\mu}$, where $\mu$ is the mean. Suppose that $|H^{ss}_{n,k}|^2$ follows an exponential distribution with mean $\mu_{|H^{ss}_{n,k}|^2}$, hence, the integrals in (\ref{integA}) and (\ref{integB}) can be simplified .
\begin{figure*}[!t]
% ensure that we have normalsize text
\normalsize
% Store the current equation number.
% Set the equation number to one less than the one
% desired for the first equation here.
% The value here will have to changed if equations
% are added or removed prior to the place these
% equations are referenced in the main text.
%\setcounter{MYtempeqncnt}{\value{equation}{17}}
\setcounter{equation}{18}
\begin{align}
&f_{\gamma_{n,k}}(\Gamma) \approx \frac{K (\sigma^2_{n} + \sigma^2_{ps}) \exp \left( -\frac{K \Gamma (\sigma^2_{n} + \sigma^2_{ps}) }{P_t \mu_{|H^{ss}_{n,k}|^2}} \right) \left(erf \left(\frac{\frac{I^{m}_{th} K}{P_t}-\mu_{N^{sp}_{m}}}{\sqrt{2 \delta^{2}_{N^{sp}_{m}}}}\right)+1\right)}{2 P_t \mu_{|H^{ss}_{n,k}|^2} } \nonumber \\ & +\frac{(\sigma^2_{n} + \sigma^2_{ps}) \delta_{N^{sp}_{m}} \exp \left( -\frac{{I^{m}_{th}}^2 K^2 \mu_{|H^{ss}_{n,k}|^2}-2 I^{m}_{th} K \mu_{N^{sp}_{m}} \mu_{|H^{ss}_{n,k}|^2} P_t+2 K (\sigma^2_{n} + \sigma^2_{ps}) P_t \delta_{N^{sp}_{m}} \Gamma +{\mu^2_{N^{sp}_{m}}} \mu_{|H^{ss}_{n,k}|^2} P^2_t}{2 \mu_{|H^{ss}_{n,k}|^2} P^2_t \delta_{N^{sp}_{m}}} \right) }{\sqrt{2 \pi} I^{m}_{th} \mu_{|H^{ss}_{n,k}|^2}} \nonumber \\ & -\frac{\begin{aligned} & 0.5 (\sigma^2_{n} + \sigma^2_{ps}) (I^{m}_{th} \mu_{N^{sp}_{m}} \mu_{|H^{ss}_{n,k}|^2}-(\sigma^2_{n} + \sigma^2_{ps}) \delta_{N^{sp}_{m}} \Gamma) \\&\times\exp \left( \frac{(\sigma^2_{n} + \sigma^2_{ps}) \Gamma ((\sigma^2_{n} + \sigma^2_{ps}) \delta_{N^{sp}_{m}} \Gamma-2 I^{m}_{th} \mu_{N^{sp}_{m}} \mu_{|H^{ss}_{n,k}|^2})}{2 {I^{m}_{th}}^2 {\mu_{|H^{ss}_{n,k}|^2}}^2} \right) \\ & \times \left(erf \left(\frac{ \left(I^{m}_{th} \mu_{|H^{ss}_{n,k}|^2} \left(\frac{I^{m}_{th} K}{P_t}-\mu_{N^{sp}_{m}}\right)+(\sigma^2_{n} + \sigma^2_{ps}) \delta_{N^{sp}_{m}} \Gamma \right)}{\sqrt{2 \delta^{2}_{N^{sp}_{m}}} I^{m}_{th} \mu_{|H^{ss}_{n,k}|^2} }\right)-1\right) \end{aligned}}{{I^{m}_{th}}^2 {\mu^{2}_{|H^{ss}_{n,k}|^2}}}
\label{PDFofgamma} 
\end{align}
% Restore the current equation number.
% IEEE uses as a separator
\hrulefill
% The spacer can be tweaked to stop underfull vboxes.
\vspace*{4pt}
\end{figure*}
%\begin{align}
%A & = \frac{1}{2} \exp \left( \frac{- K \Gamma (\sigma^2_{n} + \sigma^2_{ps})}{P_t \mu_{|H^{ss}_{n,k}|^2}} \right) \Biggr[1 + erf \biggr( \frac{\frac{I^{m}_{th} K}{P_t} - \mu_{N^{sp}_{m}}}{\sqrt{2 \delta^{2}_{N^{sp}_{m}}}} \biggr) \Biggr]
%\label{simpleA}
%\end{align}
%and
%\begin{align}
%& B = \int_{\frac{I^{m}_{th} K}{P_t}}^{\infty} \frac{\exp \left( \frac{- N^{sp}_{m} \Gamma (\sigma^2_{n} + \sigma^2_{ps})}{\mu_{|H^{ss}_{n,k}|^2} I^{m}_{th}} \right) \exp \left( \frac{- (N^{sp}_{m} - \mu_{N^{sp}_{m}})^2}{2 \delta^{2}_{N^{sp}_{m}}} \right) }{\sqrt{2 \pi \delta^{2}_{N^{sp}_{m}}}} d N^{sp}_{m} \nonumber \\ & \approx \frac{1}{2} \exp \Biggr( {\frac{\Gamma (\sigma^2_{n} + \sigma^2_{ps}) (-2 \mu_{N^{sp}_{m}} \mu_{|H^{ss}_{n,k}|^2} I^{m}_{th} + \delta^{2}_{N^{sp}_{m}} \Gamma (\sigma^2_{n} + \sigma^2_{ps}))}{2 \mu^2_{H^{ss}_{n,k}} {I^{m}_{th}}^2 }} \Biggr) \nonumber \\ & \times \Biggr[ 1 - erf \biggr( \frac{ \mu_{|H^{ss}_{n,k}|^2} I^{m}_{th} \left( - \mu_{N^{sp}_{m}} + \frac{I^{m}_{th} K}{P_{t}} \right) + \delta^{2}_{N^{sp}_{m}} \Gamma (\sigma^2_{n} + \sigma^2_{ps})}{\sqrt{2} \mu_{|H^{ss}_{n,k}|^2} I^{m}_{th} \delta_{N^{sp}_{m}}} \biggr) \Biggr].\nonumber \\
%\label{simpleB} 
%\end{align}
Finally, a closed-form expression for the pdf of $\gamma_{n,k}$ is obtained in (\ref{PDFofgamma}). 
%\begin{align}
%& F_{\gamma_{n,k}}(\Gamma) \approx \nonumber \\ & 1 - \frac{1}{2} \exp \left( \frac{- K \Gamma (\sigma^2_{n} + \sigma^2_{ps})}{P_t \mu_{|H^{ss}_{n,k}|^2}} \right) \Biggr[1 + erf \biggr( \frac{\frac{I^{m}_{th} K}{P_t} - \mu_{N^{sp}_{m}}}{\sqrt{2 \delta^{2}_{N^{sp}_{m}}}} \biggr) \Biggr] \nonumber \\ & - \frac{1}{2} \exp \Biggr( {\frac{\Gamma (\sigma^2_{n} + \sigma^2_{ps}) (-2 \mu_{N^{sp}_{m}} \mu_{|H^{ss}_{n,k}|^2} I^{m}_{th} + \delta^{2}_{N^{sp}_{m}} \Gamma (\sigma^2_{n} + \sigma^2_{ps}))}{2 \mu^2_{H^{ss}_{n,k}} {I^{m}_{th}}^2}} \Biggr) \nonumber \\ & \times \Biggr[ 1 - erf \biggr( \frac{ \mu_{|H^{ss}_{n,k}|^2} I^{m}_{th} \left( - \mu_{N^{sp}_{m}} + \frac{I^{m}_{th} K}{P_{t}} \right) + \delta^{2}_{N^{sp}_{m}} \Gamma (\sigma^2_{n} + \sigma^2_{ps})}{\sqrt{2} \mu_{|H^{ss}_{n,k}|^2} I^{m}_{th} \delta_{N^{sp}_{m}}} \biggr) \Biggr].    
%\end{align}

\subsection{Obtaining the Solutions}

It can be observed that the optimization problem, $\mathscr{O}_{1}$, is convex with respect to the transmit power $P_{n,k}(\gamma_{n,k})$. However, it is non-convex with respect to $\varphi_{n,k}(\gamma_{n,k})$ as the time-sharing factor only takes binary values. To obtain a sub-optimal solution for problem $\mathscr{O}_{1}$, we employ the Lagrangian dual decomposition algorithm. By applying dual decomposition, the non-convex optimization problem, $\mathscr{O}_{1}$, is decomposed into independent sub-problems each corresponding to a given cognitive user.

Applying the KKT conditions yields the optimal power allocation policy for Lagrangian multipliers $\mu$ and $\eta(\gamma_{n,k})$
%\begin{align}
%\frac{-\zeta  \gamma_{n,k} W\left(-\frac{\min \left(\frac{P_t}{K} , \frac{I^{m}_{th}}{N^{sp}_{m}} \right) \mu \ln(\!2\!) 2^{\frac{\text{Et} \text{Hsp} \min \left(\frac{P_t}{K} , \frac{I^{m}_{th}}{N^{sp}_{m}} \right)}{\gamma_{n,k} \text{Pdeff}}-\frac{\min \left(\frac{P_t}{K} , \frac{I^{m}_{th}}{N^{sp}_{m}} \right) \mu}{C \gamma_{n,k}}}}{\zeta  \gamma_{n,k}}\right)-\min \left(\frac{P_t}{K} , \frac{I^{m}_{th}}{N^{sp}_{m}} \right) \mu \log (2)}{\zeta  \gamma_{n,k} \mu \ln(\!2\!)}
%\end{align}
\begin{align}
&P^{*}_{n,k}(\gamma_{n,k}) = \nonumber \\&\Biggr[\frac{f_{\gamma_{n,k}}(\gamma_{n,k})}{\ln(\!2\!) (\mu f_{\gamma_{n,k}}(\gamma_{n,k})+  \eta(\gamma_{n,k}) |H^{sp}_{m,k}|^2)} - \frac{\min \left(\frac{P_t}{K} , \frac{I^{m}_{th}}{N^{sp}_{m}} \right)}{\zeta  \gamma_{n,k}} \Biggr]^{+} 
\label{Pstar}
\end{align}

%Although the optimum potential power allocation value $P^{*}_{n,k}(\gamma_{n,k})$ can be acquired in closed-form, deriving an optimal expression for the subcarrier allocation policy $\varphi^{*}_{n,k}(\gamma_{n,k})$ is more complicated. It is generally known that dual Lagrangian algorithms are challenged under linear constraints. 
%The result in (\ref{Pstar}) can be used to find the optimal subcarrier allocation strategy. By differentiating $l\Biggr(\varphi_{n,k}(\gamma_{n,k}),P_{n,k}(\gamma_{n,k}),\lambda(\gamma_{n,k}),\mu,\eta(\gamma_{n,k})\Biggr)$ with respect to $\varphi_{n,k}(\gamma_{n,k})$ we have
%\begin{align}
%& \frac{\partial l\Biggr(\varphi_{n,k}(\gamma_{n,k}),P_{n,k}(\gamma_{n,k}),\lambda(\gamma_{n,k}),\mu,\eta(\gamma_{n,k})\Biggr)}{\partial \varphi_{n,k}(\gamma_{n,k})} = \nonumber \\&\frac{\frac{\zeta  \gamma_{n,k} P^{*}_{n,k}(\gamma_{n,k})}{\min \left(\frac{P_t}{K} , \frac{I^{m}_{th}}{N^{sp}_{m}} \right)}f_{\gamma_{n,k}}(\gamma_{n,k})}{\ln(\!2\!) \left( 1 + \frac{\zeta  \gamma_{n,k} P^{*}_{n,k}(\gamma_{n,k})}{\min \left(\frac{P_t}{K} , \frac{I^{m}_{th}}{N^{sp}_{m}} \right)} \right)} \nonumber \\ & + \frac{\ln \left( 1 + \frac{\zeta  \gamma_{n,k} P^{*}_{n,k}(\gamma_{n,k})}{\min \left(\frac{P_t}{K} , \frac{I^{m}_{th}}{N^{sp}_{m}} \right)} \right)f_{\gamma_{n,k}}(\gamma_{n,k})}{\ln(\!2\!)} - \lambda_{k}(\gamma_{n,k}). 
%\label{diffVARPHI1} 
%\end{align}
%By substituting the optimal power policy (\ref{Pstar}) in (\ref{diffVARPHI1}) and by applying the KKT conditions, 
The optimal subcarrier allocation problem is 
%achieved by assigning  $\varphi_{n,k}(\gamma_{n,k}) = 0$ and $\varphi_{n,k}(\gamma_{n,k}) = 1$ where
%$\frac{\partial f_{n}}{\partial \varphi_{n,k}(\gamma_{n,k})} < 0$ and $\frac{\partial f_{n}}{\partial \varphi_{n,k}(\gamma_{n,k})} \geq 0$, respectively. The problem under consideration is
formulated as:
\begin{align}
n^{*} = argmax(\Lambda(\gamma_{n,k})) \; , \; \forall n \in \{1,...,N\} \; , \; \forall k \in \{1,...,K\}\nonumber \\
\label{PROBLEMSCAPOPT}
\end{align}
where $n^{*}$ is the optimal CRx index, and
\begin{align}
& \Lambda(\gamma_{n,k}) =  \frac{\frac{\zeta  \gamma_{n,k}{n,k} P^{*}_{n,k}(\gamma_{n,k}) }{\min \left(\frac{P_t}{K} , \frac{I^{m}_{th}}{N^{sp}_{m}} \right)}f_{\gamma_{n,k}}(\gamma_{n,k})}{\ln(\!2\!) \left( 1 + \frac{\zeta  \gamma_{n,k}{n,k} P^{*}_{n,k}(\gamma_{n,k})}{\min \left(\frac{P_t}{K} , \frac{I^{m}_{th}}{N^{sp}_{m}} \right)} \right)}\nonumber \\ & + \frac{\ln \left( 1 + \frac{\zeta  \gamma_{n,k}{n,k} P^{*}_{n,k}(\gamma_{n,k})}{\min \left(\frac{P_t}{K} , \frac{I^{m}_{th}}{N^{sp}_{m}} \right)} \right)f_{\gamma_{n,k}}(\gamma_{n,k})}{\ln(\!2\!)}.
\label{SCAPOPT}
\end{align}
where $[x]^{+} \triangleq \max\{x,0\}$. The solution in (\ref{Pstar}) can be considered as a multi-level water-filling algorithm where each subcarrier has a distinct water-level for a given user. Note that the water levels determine the potential optimum amount of power that may be allocated to $n^\text{th}$ CRx over subcarrier $k$.
The optimal subcarrier allocation policy is therefore achieved by assigning the $k^{th}$ subcarrier to the user with the highest value of $\Lambda(\gamma_{n,k})$ for all corresponding values of the $\gamma_{n,k}$. To ensure optimality, $\lambda_{k}(\gamma_{n,k})$ should be between first and second maxima of $\Lambda(\gamma_{n,k})$. If there are multiple equal maxima, the time-slot can be identically shared among the respective users. 
%Substituting (\ref{Pstar}) and (\ref{SCAPOPT}) in (\ref{LFNSFORN1}), derives $f_{n}(\varphi_{n,k}(\gamma_{n,k}),P_{n,k}(\gamma_{n,k}))$, therefore, the solution for (\ref{dualFUNCOPT1}) can be obtained. To compute the solution for the non-differentiable dual problem in    (\ref{nonDiffDual}), different optimization algorithms can be applied, including subgradient, ellipsoid, and cutting-plane. \linebreak In this work, we use the subgradient-based method to update the values of the coefficients $\lambda_{k}(\gamma_{n,k})$, $\mu$, and $\eta(\gamma_{n,k})$, in order to determine the optimal solution to (\ref{nonDiffDual}).

The subgradient method has been widely used for solving Lagrangian relaxation problems. The master problem sets the user resource allocation prices, and in order to update the dual variables, in every iteration of the subgradient method, the algorithm repeatedly finds the maximizing assignment for the sub-problems individually. For any optimal pair of $(\varphi^{*}_{n,k}(\gamma_{n,k}),P^{*}_{n,k}(\gamma_{n,k}))$, the dual variables of the problem are updated using subgradient iterations.

The potential optimum continuous-rate adaptive constellation size vector for user $n$ over subcarrier $k$ is written as
\begin{align}
& M^{*}_{n,k}(\gamma_{n,k}) = \nonumber \\ & \max \Biggl(\!1,\frac{\zeta  \gamma_{n,k}f_{\gamma_{n,k}}(\gamma_{n,k})}{\ln(\!2\!) \min\! \left(\!\frac{P_t}{K} , \frac{I^{m}_{th}}{N^{sp}_{m}} \!\right)\! (\mu f_{\gamma_{n,k}}(\!\gamma_{n,k}\!) + \eta(\!\gamma_{n,k}\!) |H^{sp}_{m,k}|^2}\! \Biggl).
\label{optimumConstellation}
\end{align}
Note that the aforementioned expression serves as an upper-bound for practical scenarios where only discrete-valued constellation sizes are applicable. Nevertheless, the real-valued $M^{*}_{n,k}(\gamma_{n,k})$ in (\ref{optimumConstellation}) may be truncated to the nearest integer. The corresponding maximum average spectral efficiency of the adaptive MQAM/OFDMA \cite{1558992} system is thus derived below
\begin{align}
& ASE^{*} = \sum_{n=1}^{N} \sum_{k=1}^{K} E_{\gamma_{n,k}} \Bigg\{  \! \log_{2} \Biggl[ \!\max \Biggl(\! 1, \nonumber \\ &  \frac{\zeta  \gamma_{n,k} f_{\gamma_{n,k}}(\gamma_{n,k})}{\ln(\!2\!)\! \min\!\! \left(\!\frac{P_t}{K} , \frac{I^{m}_{th}}{N^{sp}_{m}}\! \right)\!\! (\mu f_{\gamma_{n,k}}(\!\gamma_{n,k}\!)\! +\! \eta(\!\gamma_{n,k}\!) |H^{sp}_{m,k}|^2)}\!\! \Biggl)\!\! \Biggl] \!\varphi^{*}_{n,k}(\!\gamma_{n,k}\!)\!\! \Bigg\}.
\end{align}
%where $\gamma_{th} \geq 0$ denotes the cut-off fade depth below which the wireless channel is unused. 
According to (\ref{optimumConstellation}), no transmission takes place, i.e., $M^{*}_{n,k}(\gamma_{n,k})=1$, when $P^{*}_{n,k}(\gamma_{n,k}) = 0$. Consequently, the optimized cut-off threshold, dictated by the channel quality, power constraint, and interference constraint, is given by: $\gamma_{th} = \frac{\ln(\!2\!) (\mu + \eta(\gamma_{n,k}) |H^{m}_{sp}|^2)}{\zeta }$.

\section{Probabilistic Interference Constraint}

In a practical spectrum-sharing system, the tolerable collision level is confined by a maximum collision probability allowed by the licensed network. The tolerable collision level is highly dependent on the primary service type. For example, in case of real-time video streaming, a high collision probability is not desirable, however, delay-insensitive services can tolerate higher packet loss rates. In this section, we consider an underlay spectrum-sharing scenario where the primary users can tolerate a maximum collision probability $\varepsilon^{m}$, $\forall m \in \{1, ...,M\}$. We derive the optimal power, rate, and subcarrier allocation algorithms for the multi-user OFDMA CR system under noisy cross-link CSI availability subject to satisfying the imposed peak aggregate power and collision probability constraints. The maximization problem can be formulated as follows.

\textit{Problem} $\mathscr{O}_{2}$:
\begin{subequations}
\label{optimizationProb3}
\begin{gather}
\max_{\varphi_{n,k},P_{n,k}} \sum_{n=1}^{N} \sum_{k=1}^{K} E_{\gamma_{n,k} | \hat{h}^{sp}} \Big\{ \log_{2}(M_{n,k}(\gamma_{n,k})) \varphi_{n,k}(\gamma_{n,k}) \Big\} \label{OF3a}\\ 
\text{s. t.:} \quad \text{constraints in (\ref{OF1b}), (\ref{OF1d}), (\ref{OF1be}), and (\ref{OF1e})}, \nonumber \\
\mathscr{P} \left( \sum_{n=1}^{N} \sum_{k=1}^{K} \varphi_{n,k}(\gamma_{n,k}) P_{n,k}(\gamma_{n,k}) |H^{sp}_{m,k} | \hat{H}^{sp}_{m,k}|^2 > I^{m}_{th} \right) \leq \epsilon^{m} \nonumber \\ , \forall m \in \{1, ..., M\}.
\label{OF3c}
\end{gather}
\end{subequations}

We proceed by deriving \textit{a posteriori} distribution of the actual cross-link given the estimated channel gains. 

\textit{Proposition 1:} 
The posterior distribution of the actual channel $H^{sp}_{m,k}$ given the estimation $\hat{H}^{sp}_{m,k}$ is a complex Gaussian random variable with respective mean and variance of
\allowdisplaybreaks{
\begin{align}
& \mu_{H^{sp}_{m,k} | \hat{H}^{sp}_{m,k}} = E_{H^{sp}_{m,k} | \hat{H}^{sp}_{m,k}}(\hat{H}^{sp}_{m,k} + \Delta H^{sp}_{m,k} | \hat{H}^{sp}_{m,k}) \nonumber \\ & \hspace*{5.5em} = E_{\hat{H}^{sp}_{m,k} | \hat{H}^{sp}_{m,k}}(\hat{H}^{sp}_{m,k} | \hat{H}^{sp}_{m,k}) \nonumber \\ & + E_{\Delta H^{sp}_{m,k} | \hat{H}^{sp}_{m,k}}(\Delta H^{sp}_{m,k} | \hat{H}^{sp}_{m,k}) = (1 + \rho^2) \hat{H}^{sp}_{m,k}  
\end{align}}
and
\begin{align}
\delta^2_{H^{sp}_{m,k} | \hat{H}^{sp}_{m,k}} & = var(\hat{H}^{sp}_{m,k} + \Delta H^{sp}_{m,k} | \hat{H}^{sp}_{m,k}) \nonumber \\ & = var(\hat{H}^{sp}_{m,k} | \hat{H}^{sp}_{m,k}) + var(\Delta H^{sp}_{m,k} | \hat{H}^{sp}_{m,k}) \nonumber \\ & + 2 cov(\Delta H^{sp}_{m,k} | \hat{H}^{sp}_{m,k},\hat{H}^{sp}_{m,k} | \hat{H}^{sp}_{m,k}) \nonumber \\ & = (1 - \rho^2) \delta^{2}_{\Delta H^{sp}_{m,k}}.
\end{align}

%\begin{figure*}[ht]
%\begin{minipage}[b]{0.49\linewidth}
%\centering
%\includegraphics[width=\textwidth]{Chi_2_2_to_New_Chi.eps}
%\subcaption{$\beta^{m}_{k} \sim \text{Chi-Square}(2,2)$, $\delta^{2}_{H^{sp}_{m,k} | \hat{H}^{sp}_{m,k}}=1$.}
%\label{fig:figure1}
%\end{minipage}
%\hfill
%\begin{minipage}[b]{0.49\linewidth}
%\centering
%\includegraphics[width=\textwidth]{Gamma_2_5_4_to_New_Chi_VarNim.eps}
%\subcaption{$\beta^{m}_{k} \sim \text{Gamma}(2,0.5,4)$, $\delta^{2}_{H^{sp}_{m,k} | \hat{H}^{sp}_{m,k}}=0.5$.}
%\label{fig:figure2}
%\end{minipage}
%\centering{}\caption{Approximated Model and Empirical Data cdfs, obtained from Monte-Carlo simulations.
%\label{cdf-1-1}}
%\end{figure*}
%
%\begin{figure*}
%\centering
%\subfloat[$\beta^{m}_{k} \sim \text{Chi-Square}(2,2)$, $\sigma^{2}_{H^{sp}_{m,k} | \hat{H}^{sp}_{m,k}}=1$.]{
%\includegraphics[scale=0.18]{Chi_2_2_to_New_Chi.eps}
%}
%\subfloat[$\beta^{m}_{k} \sim \text{Gamma}(2,0.5,4)$, $\sigma^{2}_{H^{sp}_{m,k} | \hat{H}^{sp}_{m,k}}=0.5$.]{
%\includegraphics[scale=0.18]{Gamma_2_5_4_to_New_Chi_VarNim.eps}
%}
%\centering{}\caption{Approximated Model and Empirical Data cdfs, obtained from Monte-Carlo simulations.
%\label{cdf-1-1}}
%\end{figure*}

Assuming equal variance $\delta^2_{H^{sp}_{m,k} | \hat{H}^{sp}_{m,k}}$ across all users and subcarriers, the collision probability constraint in (\ref{OF3c}) can be expressed as
\begin{align}
\mathscr{P}\!\! \left(\!\!\delta^2_{H^{sp}_{m,k} | \hat{H}^{sp}_{m,k}} \sum_{n=1}^{N} \!\! \sum_{k=1}^{K} \! \varphi_{n,k}(\!\gamma_{n,k}\!) P_{n,k}(\!\gamma_{n,k}\!) |\Xi^{m}[k]|^2\! >\! I^{m}_{th} \!\! \right)\! \! \leq \!\! \varepsilon^{m}
\label{CHAINOTNOT}
\end{align}
where $\Xi^{m}[k]$ is a complex Gaussian random variable with variance of one and mean of
\begin{align}
\mu_{\Xi^{m}[k]} = \abs[\Bigg]{ \frac{\mu_{H^{sp}_{m,k} | \hat{H}^{sp}_{m,k}}}{\delta_{H^{sp}_{m,k} | \hat{H}^{sp}_{m,k}}} } ^2.
\end{align}

It should be noted that in contrast to the sum of equally-weighted Chi-Square random variables in Lemma 1, (\ref{CHAINOTNOT}) includes a sum of non-equally-weighted Chi-Square random variables. 
In general, obtaining the exact distribution of the linear combination of weighted Chi-Square random variables is rather complex. Although several approximations have been proposed in the literature, e.g., \cite{springerlink:10.1007/BF02932611,springerlink:10.1007/BF02595410,1987}, most are not easy to implement. In this work, similar to \cite{7004853} we use the following approximation:
\begin{align}
\delta^{'} = \sum_{k=1}^{K} \mu_{\Xi^{m}[k]} \label{APsub1}\\
D = 2K \label{APsub2}\\
\xi = \frac{\sum_{k=1}^{K} \beta^{m}_{k} (2 + \mu_{\Xi^{m}[k]})}{2K + \sum_{k=1}^{K} \mu_{\Xi^{m}[k]}}. \label{APsub3}
\end{align}

 where $\delta^{'}$, $D$, and $\xi$ are respectively the non-centrality parameter, degree of freedom, and weight of the new random variable.
Now (\ref{CHAINOTNOT}) can be simplified to:
\begin{align}
& \mathscr{P}\!\! \Biggr( \!\delta^2_{H^{sp}_{m,k} | \hat{H}^{sp}_{m,k}} \sum_{n=1}^{N} \sum_{k=1}^{K} \varphi_{n,k}(\gamma_{n,k}) P_{n,k}(\gamma_{n,k}) |\Xi^{m}[k]|^2 > I^{m}_{th}\! \Biggr) \nonumber \\ & \approx Pr(\xi \chi_{D}^{2}(\delta^{'}) > I^{m}_{th}).
\end{align}
According to \cite{1987}, since the non-centrality parameter is small relative to the degree of freedom, we can approximate the non-central Chi-Square distribution with a central one using the following
\begin{align}
\mathscr{P} (\xi \chi_{D}^{2}(\delta^{'}) > I^{m}_{th}) \approx \mathscr{P} (\chi_{D}^{2}(0) > \frac{I^{m}_{th}/\xi}{1 + \delta^{'}/D}).
\label{PrNCPrC}
\end{align}
Then similar to \cite{7004853} the deterministic inequality 
\begin{gather}
\delta^2_{H^{sp}_{m,k} | \hat{H}^{sp}_{m,k}} \sum_{k=1}^{K} (2 + \mu_{\Xi^{m}[k]}) \sum_{n=1}^{N} \varphi_{n,k}(\gamma_{n,k}) P_{n,k}(\gamma_{n,k}) \nonumber \\ \leq \frac{K \, I^{m}_{th}}{(K!)^{1/K} \ln \left( 1 - (1 - \varepsilon^{m})^{1/K} \right)} 
\label{DetermIC}
\end{gather}
satisfies the probabilistic inequality (\ref{CHAINOTNOT}). Therefore, the constraint (\ref{CHAINOTNOT}) can be replaced by (\ref{DetermIC}).

To obtain $ASE^{*}$ for the probabilistic interference constraint and `probabilistic case' of estimation error scenario, we employ the Lagrangian dual optimization method as in the previous sections, where
\begin{align}
\alpha_{k} = \delta^2_{H^{sp}_{m,k} | \hat{H}^{sp}_{m,k}} (2 + \mu_{\Xi^{m}[k]})
\end{align}
and 
\begin{align}
\overline{I^{m}_{th}} = \frac{K \, I^{m}_{th}}{(K!)^{1/K} \ln \left( 1 - (1 - \varepsilon^{m})^{1/K} \right)}.
\end{align}
Therefore, by solving the Lagrangian optimization problem the following optimal power allocation solution can be obtained for user $n$ over subcarrier $k$
\begin{align}
&P^{*}_{n,k}(\gamma_{n,k}) =\nonumber \\ & \Biggr[\frac{f_{\gamma_{n,k}}(\gamma_{n,k})}{\ln(\!2\!) (\mu f_{\gamma_{n,k}}(\gamma_{n,k})\! + \!\eta(\gamma_{n,k}) \alpha_{k})} - \frac{{\min \left(\frac{P_t}{K} , \frac{\overline{I^{m}_{th}}}{\hat{N}^{sp}_{m}} \right)}}{\zeta  \gamma_{n,k}} \Biggr]^{+} 
\label{Pstar2}
\end{align}
where $\hat{N}^{sp}_{m}$ in the `probabilistic case' is derived in Appendix A, Section C.
The optimal subcarrier allocation policy is the solution to the following problem
\begin{align}
n^{*} = argmax(\Lambda(\gamma_{n,k})) \; , \; \forall n \in \{1,...,N\} \; , \; \forall k \in \{1,...,K\}
\label{PROBLEMSCAPOPT4}
\end{align}
where $n^{*}$ is the optimal CRx index, and
\begin{align}
& \Lambda(\gamma_{n,k}) =  \frac{\frac{\zeta  \gamma_{n,k} P^{*}_{n,k}(\gamma_{n,k})}{{\min \left(\frac{P_t}{K} , \frac{\overline{I^{m}_{th}}}{\hat{N}^{sp}_{m}} \right)}}f_{\gamma_{n,k}}(\gamma_{n,k})}{\ln(\!2\!) \left( 1 + \frac{\zeta  \gamma_{n,k} P^{*}_{n,k}(\gamma_{n,k})}{{\min \left(\frac{P_t}{K} , \frac{\overline{I^{m}_{th}}}{\hat{N}^{sp}_{m}} \right)}} \right)}\nonumber \\ & + \frac{\ln \left( 1 + \frac{\zeta  \gamma_{n,k} P^{*}_{n,k}(\gamma_{n,k})}{{\min \left(\frac{P_t}{K} , \frac{\overline{I^{m}_{th}}}{\hat{N}^{sp}_{m}} \right)}} \right)f_{\gamma_{n,k}}(\gamma_{n,k})}{\ln(\!2\!)}.
\label{SCAPOPT4}
\end{align}
By employing the sub-gradient method, optimal expressions are derived for the constellation size and hence spectral efficiency under collision probability constraint and imperfect cross-link CSI: 

%the Lagrangian multipliers $\mu$ and $\eta(\gamma_{n,k})$ can be updated by 
%\begin{gather}
%\mu^{i+1} = \mu^{i} - \tau^{i}_{1} \biggr( P_{t} - \sum_{n=1}^{N} \sum_{k=1}^{K} \varphi^{*}_{n,k}(\gamma_{n,k}) P^{*}_{n,k}(\gamma_{n,k}) \biggr) \label{SUB14} \\
%\eta^{i+1}(\gamma_{n,k}) = \eta^{i}(\gamma_{n,k}) - \tau^{i}_{2} \nonumber \\ \times \! \left(\! \overline{I^{m}_{th}} \! - \! \delta^2_{H^{sp}_{m,k} | \hat{H}^{sp}_{m,k}}\! \sum_{k=1}^{K} (2 \! + \! \mu_{\Xi[k]}) \sum_{n=1}^{N} \varphi^{*}(\gamma_{n,k}) P^{*}_{n,k}(\gamma_{n,k})\! \right). 
%\label{SUB24}
%\end{gather}

\begin{align}
&M^{*}_{n,k}(\gamma_{n,k}) = \nonumber \\ &\max \Biggl(1,\frac{\zeta  \gamma_{n,k}f_{\gamma_{n,k}}(\gamma_{n,k})}{\ln(\!2\!) {\min \left(\frac{P_t}{K} , \frac{\overline{I^{m}_{th}}}{\hat{N}^{sp}_{m}} \right)} (\mu f_{\gamma_{n,k}}(\gamma_{n,k}) + \eta(\gamma_{n,k}) \alpha_{k})} \Biggl),
\label{optimumConstellation4}
\end{align}
%\begin{align}
%& AASE^{*} = \nonumber \\ & \sum_{n=1}^{N} \sum_{k=1}^{K} \! \int_{\gamma_{th}}^{\infty} \! \log_{2} \! 
%\Bigg[ 
%\max \Biggl(1,\frac{\zeta  |H^{ss}_{n,k}|^2}{\ln(\!2\!) ({\min \left(\frac{P_t}{K} , \frac{I^{m}_{th}}{N^{sp}_{m}} \right)}) (\mu + \eta(\gamma_{n,k}) \alpha_{k})} \Biggl) 
%\Bigg]
%\nonumber \\ & \times \varphi^{*}[n,k] p_{\gamma_{n,k} | \hat{H}^{sp}_{m,k}}(\gamma_{n,k} | \hat{H}^{sp}_{m,k}) \, d\gamma_{n,k}. 
%\end{align}
\begin{align}
& ASE^{*} = \sum_{n=1}^{N} \sum_{k=1}^{K} \sum_{\gamma_{n,k} | \hat{h}^{n,k}} \Bigg\{   \log_{2} \Biggl[ \max \Biggl( 1, \nonumber \\ &  \frac{\zeta  \gamma_{n,k} f_{\gamma_{n,k}}(\gamma_{n,k})}{\ln(\!2\!) \min \left(\!\frac{P_t}{K} , \frac{\overline{I^{m}_{th}}}{\hat{N}^{sp}_{m}}\! \right)\!\! (\mu f_{\gamma_{n,k}}(\gamma_{n,k})\! + \!\eta(\gamma_{n,k}) \alpha_{k})} \!\!\Biggl) \!\Biggl]\! \varphi^{*}_{n,k}(\!\gamma_{n,k}\!)\! \Bigg\}.
\end{align}

%The methodologies for deriving the expressions of the cdf of the received SINR given the estimation, for different `average case', `worst case', and `probabilistic case' scenarios of estimation error, are elucidated in Appendix B. 

\section{Discussion of Results}

In this section, we examine the performance of an OFDMA CR network operating under total average transmit power and deterministic/probabilistic peak aggregate interference constraints with perfect/imperfect cross-channel estimation using the respective optimal resource allocation solutions. In the following results, perfect CSI knowledge of the cognitive user link is assumed to be available at the CTx through an error-free feedback channel. Thus, $|H^{ss}_{n,k}|$, $\forall \{n,k\}$, are drawn through a Rayleigh distribution. Furthermore, the secondary-secondary power gain mean gains, $\mu_{|H^{ss}_{n,k}|^2}$, $\forall \{n,k\}$, are taken as Uniformly-distributed random variables within 0 to 2. It should be noted that the sub-channels are assumed to be narrow-enough so that they experience frequency-flat fading. 
%Two scenarios are considered for cross-link channel gains, $H^{sp}_{m,k}$, $\forall \{m,k\}$: perfect and noisy. 
Interfering cross-channel values, $H^{sp}_{m,k}$, $\forall \{m,k\}$, are distributed according to a complex Gaussian distribution with mean 0.05 and variance 0.1. For the inaccurate cross-link CSI case, the channel estimation and error for all sub-channels are taken as independent and identically distributed (i.i.d.) zero-mean Normally-distributed random variables. In addition, the AWGN power spectral density is set to -174 dBm. The total average power constraint is imposed on the system in all cases. 
%To evaluate the performance of our proposed adaptive MQAM/OFDMA system, we also introduce a non-optimized scheme, where the maximum non-adaptive constellation size, i.e., $M_{n,k}(\Upsilon) =  M_{n,k}$, $\forall \{n,k\}$, is derived based on the target-BER constraint. 
Discrete-rate cases with real-valued MQAM signal constellations, i.e., $\log_{2}(M) \in \{2,4,6,8,10\}$ bits/symbol, are also considered for practical scenarios. All results correspond to the scenario with three cognitive receivers and a single primary receiver, hence, the subscript $m$ is hereafter omitted. 

\begin{figure}[t]
\centering
\includegraphics[width=.5\textwidth]{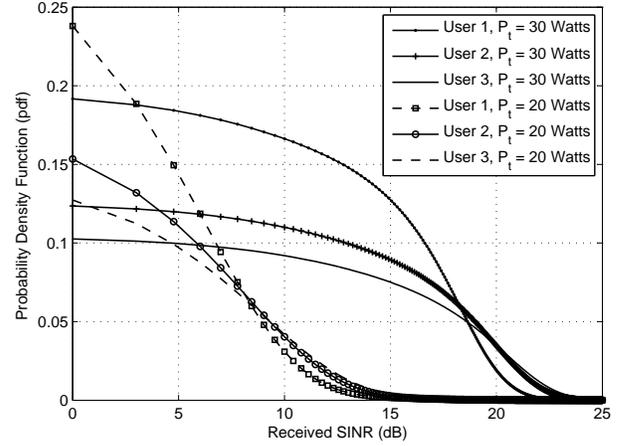} 
\caption{Probability density functions of the received SINR for OFDMA users in a given subcarrier $k$ under different average power constraint values. System parameters are: $K = 64$, $k = 16$, $I_{th} = 5$ Watts.}
\label{fig:PDFvsSINR}
\end{figure}

\begin{figure}[t]
\centering
\includegraphics[width=.5\textwidth]{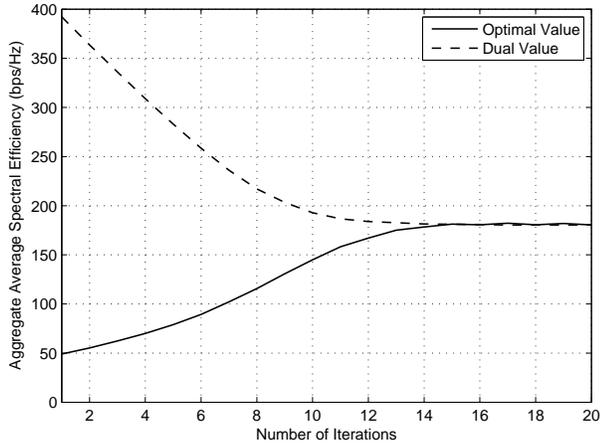} 
\caption{Optimal and dual values versus the number of iterations using the sub-gradient method. Results for the case with deterministic interference constraint and perfect cross-link CSI knowledge. System parameters are: $K = 64$, $P_t = 30$ Watts, $I_{th} = 10$ Watts, $\xi = 10^{-2}$.}
\label{fig:subgrad}
\end{figure}

The approximated probability distributions of the received SINRs for cognitive users in a randomly taken subcarrier, i.e., here $k=16$, under different total average power constraint limits $P_t$ is plotted in Fig. \ref{fig:PDFvsSINR}. For a fixed interference constraint of $I_{th} = 5$ Watts, it can be observed that the probability of higher received SINR improves as the value of $P_t$ increases. For example, for user 3, the probability of receiving $\gamma_{3,16} = 10$ dB is 54.5\% higher as the value of $P_t$ is increased from 20 to 30 Watts. 

Fig. \ref{fig:subgrad} illustrates the evolution of the optimal and dual values using the sub-gradient method over time. The results correspond to the maximum deliverable ASE for the case with deterministic interference constraint and perfect cross-link CSI knowledge. The iterative sub-gradient algorithm converges quickly and typically achieves a lower-bound at 96.5\% of the optimal value within 12 iterations. It can easily be shown that the proposed dual decomposition algorithm converges fast for different parameters of system settings. 

\begin{figure}[t]
\centering
\includegraphics[width=.5\textwidth]{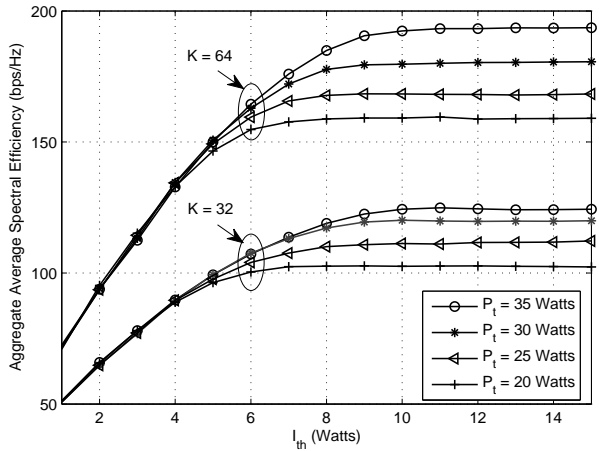} 
\caption{ASE performance versus the tolerable interference power threshold level with different values of $P_t$ and $K$. Results for the case with deterministic interference constraint and perfect cross-link CSI knowledge. System parameters are: $I_{th} = 10$ Watts, $\xi = 10^{-2}$.}
\label{fig:ASEvsPt}
\end{figure}

\begin{figure}[t]
\centering
\includegraphics[width=.5\textwidth]{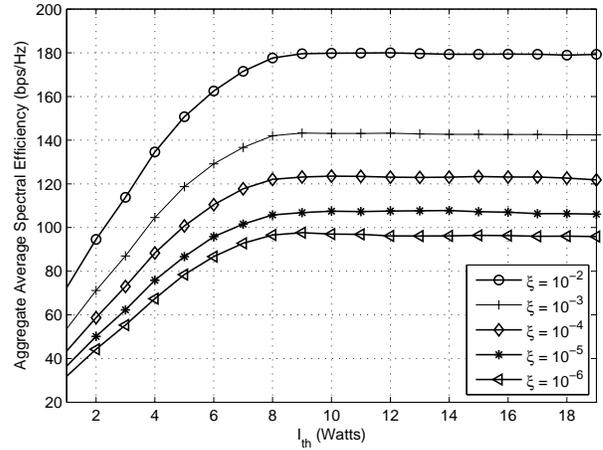} 
\caption{ASE performance using the proposed RRA algorithm versus $I_{th}$ constraint for different BER-target values. Results correspond to the case with deterministic interference constraint and perfect cross-link CSI. System parameters are: $K = 64$, $P_t = 30$ Watts.}
\label{fig:ASEvsBER}
\end{figure}

%\begin{figure}[t]
%\centering
%\includegraphics[width=.5\textwidth]{ASE_Rho_Pt_AC.eps} 
%\caption{Achievable ASE with imperfect cross-link CSI and `average case' of estimation error against $\rho$ for different values of $P_t$. System parameters are: $K = 64$, $I_{th} = 25$ Watts, $\xi = 10^{-2}$, $\delta^{2}_{\hat{H}^{sp}_{k}} = 1$.}
%\label{fig:ASEaveragecasevsRho}
%\end{figure}

%\begin{figure}[t]
%\centering
%\includegraphics[width=.5\textwidth]{ASE_Rho_Pr_WC.eps} 
%\caption{Achievable ASE with imperfect cross-link CSI and `worst case' of estimation error against $\rho$ with $pr$. System parameters are: $K = 64$, $P_t = 20$ Watts, $I_{th} = 5$ Watts, $\xi = 10^{-3}$, $\delta^{2}_{\hat{H}^{sp}_{k}} = 1$.}
%\label{fig:ASEworstcasevsRho}
%\end{figure}

\begin{figure}[t]
\centering
\includegraphics[width=.5\textwidth]{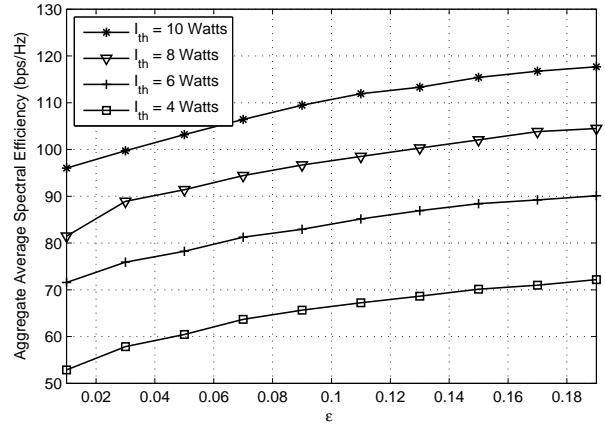} 
\caption{Achievable ASE with imperfect cross-link CSI and `probabilistic case' of estimation error against $\epsilon$ with $I_{th}$. System parameters are: $K = 64$, $P_t = 40$ Watts, $\xi = 10^{-3}$, $\rho = 0.5$, $\delta^{2}_{\hat{H}^{sp}_{k}} = 1$.}
\label{fig:ASEprobcasevsEps}
\end{figure}

%\begin{figure}[t]
%\centering
%\includegraphics[width=.5\textwidth]{ASE_Compare_Ith.eps} 
%\caption{Performance under imperfect cross-link CSI for different cases of estimation error against $I_{th}$. System parameters are: $K = 64$, $P_t = 45$ Watts, $\xi = 10^{-2}$, $pr = 0.95$, $\rho = 0.2$, $\epsilon = 5\%$, $\delta^{2}_{\hat{H}^{sp}_{k}} = 0.1$.}
%\label{fig:ASEcompareIth}
%\end{figure}

Fig. \ref{fig:ASEvsPt} shows the achievable ASE of the adaptive MQAM/OFDMA CR system versus CTx-PRx interference power threshold levels under total average power and deterministic interference constraints with perfect cross-link CSI knowledge. As expected, greater ASE values are achieved for higher maximum tolerable interference since $I_{th}$ limits the cognitive users' transmit power. The improved performance however approaches a plateau in the high $I_{th}$ region as the $P_t$ threshold becomes the dominant power constraint. Note that the improved performance by increasing $I_{th}$ is obtained at the cost of increased probability of violating the primary users' QoS. Further, imposing a higher maximum peak average power setting enhances the achievable ASE in high $I_{th}$ region - $P_t$, for the particular values taken in this example, achieve the same ASE over small $I_{th}$ settings. Moreover, increasing the number of subcarriers results in higher attainable performance.    

Achievable ASE performance under different maximum tolerable interference thresholds for respective values of BER-target with perfect cross-link CSI availability is shown in Fig. \ref{fig:ASEvsBER}. It can be seen that the system performance is improved under less stringent QoS constraints. For example, a 26.9\% gain in ASE performance is achieved by imposing $\xi = 10^{-2}$ in comparison to $\xi = 10^{-3}$. However, the gap in performance becomes less significant for lower BER-target regimes. 

%System performance with noisy cross-link CSI and `average case' of estimation error versus the correlation factor between estimation and error variables, $\rho$, is depicted in Fig. \ref{fig:ASEaveragecasevsRho}. It can be seen that a higher correlation factor increases the likelihood between true and estimated interfering channels, hence, the probability of violating the interference constraint on average is improved and in turn a lower ASE for the cognitive system is realized.  
%%Talk about how changing rho is effectively changing the error variance. 
%%The achievable ASE with imperfect cross-link CSI knowledge and `worst case' of estimation error against $\rho$ for different probabilities of channel estimation error bound $pr$ is studied in Fig. \ref{fig:ASEworstcasevsRho}.
Apart from the effect of $\rho$ on the performance, higher values of $pr$ increase the robustness of the interference management scheme but come at the cost of lower achievable spectral efficiencies. The results indicate that the improved ASE performance by decreasing $pr$ in the lower half region (i.e., $pr \leq 0.5$) is not significant yet it may cause critical interference to the primary service operation. For example, given $\rho = 0.5$, varying the value of $pr$ from 0.5 to 0.1 results in a 40\% increase in the probability of error bound violation but only provides an effective gain of 2.3\% in the cognitive system performance.      

The achievable performance with imperfect cross-channel information and `probabilistic case' of estimation error versus the collision probability $\epsilon$ with respective $I_{th}$ values is illustrated in Fig. \ref{fig:ASEprobcasevsEps}. Increasing the maximum probability of violating the interference constraint significantly improves the spectral efficiency of the cognitive network. The tradeoff is however the degradation of the primary service operation which is deemed highly undesirable in practical scenarios.   

%System performance with noisy cross-link CSI for the `average case', `worst case', and `probabilistic case' of estimation error is demonstrated in Fig. \ref{fig:ASEcompareIth}. The results show that the `probabilistic case' with $5\%$ collision probability outperforms the achievable ASE under the `worst case' scenario with an error bound of $pr = 0.5$. For example, given $I_{th} = 6$ Watts, the `probabilistic case' achieves a 26.7\% gain in ASE over the `worst case'. Moreover, employing the `average case' provides higher spectral efficiencies. For instance, a 7.0\% increase in performance is achieved utilizing the `average case' over the `probabilistic case'. For high values of $I_{th}$, the total average power constraint becomes the dominant limit and therefore the performance under different cases of estimation error eventually converge. Note that the `average case' controls the interference based on the average error estimation, hence, it cannot mitigate the potential instantaneous interference violations. On the other hand, implementing the `worst case' can guarantee that the interference constraints are adhered to at any given time, thus, preserving the primary users' QoS. The proposed `probabilistic case' of estimation error provides an optimal trade-off  between the achievable performance of cognitive system and managing the QoS of primary users. In particular, the `probabilistic case' is advantageous in terms of performance and flexibility over the conventional `average case' and `worst case' scenarios. 

\section{conclusions}

In this paper, we have studied the spectral efficiency performance of adaptive MQAM/OFDMA underlay CR networks with certain/uncertain interfering channel information. We derived novel RRA algorithms to enhance the overall cognitive system performance subject to satisfying total average power and peak aggregate interference constraints. The proposed framework considers  both cases of perfect and imperfect cross-link CSI knowledge at the cognitive transmitter. To compute the average spectral efficiency, we developed mathematical close form for the distribution of the received SINR for given users over different sub-channels in the respective cases under consideration. Through simulation results we studied the achievable performance of the cognitive system using our proposed RRA algorithms. By adapting the power, rate, and subcarrier allocation policies to the time-varying secondary-secondary fading channels and secondary-primary interfering channels, a significant gain in the spectral efficiency performance of the cognitive system can be realized, whilst controlling the interference on the primary service receivers. Furthermore, the impact of parameters uncertainty on overall system performance was investigated. The proposed `probabilistic case' in this paper, which was derived as a low complexity deterministic constraint, provided an optimal trade-off between the achievable performance of the cognitive network and preserving the QoS of the primary users. 

\bibliographystyle{IEEEtran}
{\footnotesize
\bibliography{IEEEabrv,myref}}

\end{document}